\documentclass[12pt]{article}
\usepackage{graphics}
\usepackage{amssymb}

\newcommand{\ie}{{\em i.e.}}
\newcommand{\eg}{{\em e.g.}}
\newcommand{\QED}{\mbox{\rule[-1.5pt]{6pt}{10pt}}}
\newcommand{\lhs}{l.h.s. }
\newcommand{\rhs}{r.h.s. }
\newcommand{\im}{{\rm Im\,}}
\newcommand{\Ker}{{\rm Ker\,}}
\newcommand{\D}{{\rm d}}
\newcommand{\e}{{\rm e}}

\newcommand{\C}{\mathbb{C}}
\newcommand{\R}{\mathbb{R}}
\newcommand{\Z}{\mathbb{Z}}
\newcommand{\BB}{{\cal B}}
\newcommand{\DD}{{\cal D}}

\newtheorem{claim}{Claim}[section]
\newtheorem{theorem}[claim]{Theorem}
\newtheorem{proposition}[claim]{Proposition}
\newtheorem{lemma}[claim]{Lemma}

\newtheorem{example}[claim]{Example}

\begin{document}

\title{Bound states of infinite curved polymer chains}
\author{P.~Exner}
\date{}
\maketitle
\begin{quote}
{\small \em Department of Theoretical Physics, Nuclear Physics
Institute, \\ \phantom{e)x}Academy of Sciences, CZ-25068 \v Re\v z, and \\
 Doppler Institute, Czech Technical University, B\v{r}ehov{\'a} 7,\\
\phantom{e)x}CZ-11519 Prague, Czech Republic \\
 \rm \phantom{e)x}exner@ujf.cas.cz}
\vspace{8mm}

\noindent {\small We investigate an infinite array of point
interactions of the same strength in $\R^d,\, d=2,3$, situated at
vertices of a polygonal curve with a fixed edge length. We
demonstrate that if the curve is not a line, but it is
asymptotically straight in a suitable sense, the corresponding
Hamiltonian has bound states. Example is given in which the number
of these bound states can exceed any positive integer.}
\end{quote}


\section{Introduction}

Methods of guiding particles or light quanta along a prescribed
path are of interest from both theoretical and practical point of
view. One aspect of this problem are relations between the
geometry of the ``channel'' and the spectral and scattering
properties of the corresponding Hamiltonian. The last decade
brought various results in this field -- see, \eg, the paper
\cite{DE} and the recent books \cite{Hu, LCM} for an extensive
bibliography.

A particular question to be addressed in this letter concerns the
existence of curvature-induced bound states in infinite channels
which are in some sense asymptotically straight. This effect was
first demonstrated in \cite{ES} and subsequently studied by
numerous authors. A common feature of these studies, however, is
that they assume a strict localization in the sense that the
configuration space is a neighborhood of a given curve. This is
not fully realistic, because in actual ``quantum waveguides'' the
confinement comes from a potential well of a finite depth, and it
is not a priori clear, whether the binding effect will persist in
the presence of a tunneling between different parts of the
channel.

The aim of the present letter is to give an affirmative answer to
the above question in a well-known model of a ``polymer chain'',
\ie, an array of point interactions in $\R^d,\, d=2,3$, with fixed
coupling parameter and the distance between the neighbors, which
is certainly a very weak way to keep the particle ``within'' the
channel. The straight-polymer spectrum is thoroughly analyzed in
\cite{AGHH}: it is purely absolutely continuous and below bounded
with at most one gap. We will show that making the chain curved --
locally in a suitable sense -- will lead to emergence of isolated
eigenvalues below the continuum threshold, and that making the
curvature ``large'' enough we can produce many such bound states.


\setcounter{equation}{0}
\section{Formulation of the problem} \label{formul}

Let $Y=\{y_n\}_{n\in\Z}$ be a sequence in $\R^d,\; d=2,3$, with the
following property: there is an $\ell>0$ such that
\begin{equation} \label{geobound}
|y_j-y_{j'}| \le \ell\,|j-j'|
\end{equation}
holds for any integers $j,j'$. In particular, the distance between
neighboring points satisfies
\begin{equation} \label{neighbor}
|y_j-y_{j+1}| =\ell
\end{equation}
for each $j\in\Z$. For simplicity we shall use the symbol $Y$ both
for a map $\Z\to\R^d$ and the subset of $\R^d$ which is the range of this
map. Furthermore, we shall assume:
\begin{description}
 \vspace{-1.2ex}
 \item{\em (a1)} there is $c_1\in(0,1)$ such that $|y_j-y_{j'}| \ge
 c_1\ell\,|j-j'|$. In particular, $Y$ as a map is injective, and if it
 has asymptotes for $j\to \pm\infty$ they are not parallel.
 \vspace{-1.2ex}
 \item{\em (a2)} $\:Y$ is asymptotically straight in the
 following sense: there are positive $c_2,\, \mu$, and $\omega\in(0,1)$
 such that the inequality
\begin{equation} \label{asympt}
1-\, {|y_j-y_{j'}|\over|j-j'|} \le c_2 \left\lbrack
1+|j+j'|^{2\mu} \right\rbrack^{-1/2}
\end{equation}
 holds in the sector $S_\omega:= \left\{ (j,j'):\: j,j'\ne0,\,
 \omega < {j\over j'} < \omega^{-1}\, \right\}$ of $\Z^2$.
\end{description}
The operators we shall investigate are the point-interaction
Hamiltonians $H_{\alpha,Y} \equiv -\Delta_{\alpha,Y}$ with the
same interaction ``strength'' at each point, which in the notation
of \cite{AGHH} means that $\alpha$ is a constant sequence. They
are defined by means of the boundary conditions
\begin{equation} \label{bc}
L_1(\psi,y_j) -\alpha L_0(\psi,y_j)=0\,, \quad j\in\Z\,,
\end{equation}
expressed in terms of the generalized boundary values
$$ L_0(\psi,y):= \lim_{|x-y|\to 0}\, {\psi(x)\over
\phi_d(x\!-\!y)}\,, \; L_1(\psi,y):= \lim_{|x-y|\to 0}
\bigl\lbrack \psi(x)- L_0(\psi,y)\, \phi_d(x-y) \bigr\rbrack\,, $$
where $g_d$ are the appropriate fundamental solutions,
$$ \phi_2(x)= -{1\over2\pi}\, \ln|x|\,, \quad \phi_3(x) = {1\over
4\pi|x|}\,, $$
related to the free Green's functions
\begin{equation} \label{freeG}
G_k(x\!-\!x') = \left\lbrace \begin{array}{lcc} {i\over 4}
H_0^{(1)}(k|x-x'|) & \quad \dots \quad & d=2 \\
{\e^{ik|x-x'|}\over 4\pi |x-x'|} & \quad \dots \quad & d=3
\end{array} \right.
\end{equation}
More exactly, one defines in this way the point-interaction
Hamiltonian $H_{\alpha,\tilde Y}$ for any finite subset $\tilde
Y\subset Y$, and $H_{\alpha,Y}$ is obtained as the strong
resolvent limit over the filter of finite subsets
\cite[Secs.~III.1, III.4]{AGHH}. The resolvent of $H_{\alpha,Y}$
is given Krein's formula,
\begin{equation} \label{krein}
(-H_{\alpha,Y}-k^2)^{-1} = G_k + \sum_{j,j'=1}^{\infty}
[\Gamma_{\alpha,Y}(k)]^{-1}_{jj'} \left(
\overline{G_k(\cdot\!-\!y_{j'})}, \cdot \right) G_k(\cdot\!-\!y_j)
\end{equation}
for $k^2\in\rho\left( H_{\alpha,Y}\right)$ with $\im k>0$, where
$\Gamma_{\alpha,Y}(k)$ is a closed operator (which is bounded in
our case) on $\ell^2(\Z)$ the matrix representation of which is
\begin{equation} \label{Gamma}
\Gamma_{\alpha,Y}(k):= \bigg\lbrack (\alpha-\xi_d^k) \delta_{jj'}
- \tilde g^{Y,k}_{jj'} \bigg\rbrack_{j,j'\in\Z}\,,
\end{equation}
where $\xi^k_d$ is the regularized Greens's function
$$ \xi^k_2= -{1\over 2\pi}\, \left( \ln{k\over 2i} + \gamma
\right)\,, \quad \xi^k_3= {ik \over 4\pi}\,, $$
with $\gamma=-\psi(1)$ the Euler number, and
$$ \tilde g^{Y,k}_{jj'} = \left\lbrace \begin{array}{lcc}
G_k(y_j-y_{j'}) & \quad \dots \quad & j\ne j' \\ 0 & \quad \dots
\quad & j=j' \end{array} \right. $$
Since $\alpha$ is independent of $j$, the map $k\mapsto
\Gamma_{\alpha,Y}(k)$ is analytic in the open upper halfplane.
Moreover, $\Gamma_{\alpha,Y}(k)$ is boundedly invertible for $\im
k>0$ large enough, while for $k\in\C_+$ not too far from the real
axis it may have a nontrivial null-space. By (\ref{krein}) the
latter determines the spectrum of the original operator
$H_{\alpha,Y}$ on the negative halfline. In particular, one easily
checks the following result.
\begin{lemma} \label{BS}
(i) A point $-\kappa^2<0$ belongs to $\rho\left(
H_{\alpha,Y}\right)$ iff $\,\Ker \Gamma_{\alpha,Y}= \{0\}$. \\
(ii) If the operator-valued function $\kappa\mapsto
\Gamma_{\alpha,Y} (i\kappa)^{-1}$ has bounded values in an open
interval $I\subset\R_+$ with the exception of a point $\kappa_0\in
I$, where $\dim\Ker \Gamma_{\alpha,Y}(i\kappa)=n$, then
$-\kappa_0^2$ is an isolated eigenvalue of $H_{\alpha,Y}$ of
multiplicity $n$.
\end{lemma}


\setcounter{equation}{0}
\section{The essential spectrum} \label{ess-sec}

Let us now ask how the geometry of the array $Y$ is reflected in
the spectral properties of the operator $H_{\alpha,Y}$. As a
departure point we remind some facts about straight polymers. If
we denote by $Y_0$ the linearly arranged sequence, $|y_j-y_{j'}| =
\ell\,|j-j'|$ for all $j,j'\in\Z$, the corresponding spectrum is
purely absolutely continuous and consists of two bands -- see
\cite[Secs.~III.1.5. III.4]{AGHH} -- which may overlap if $\alpha$
is not large enough negative, in particular, for $d=3$ and
$\alpha\ell \ge -{1\over 2\pi} \ln2$. Its threshold
$E_d^{\alpha,\ell} \equiv E_d^{\alpha,\ell}(0)$ is always
negative. In the three-dimensional case it is known explicitly,
\begin{equation} \label{thres3}
E_3^{\alpha,\ell} = {1\over \ell^2} \left\lbrack
\ln\left(1+{1\over 2}\, \e^{-4\pi\alpha\ell} +
\e^{-2\pi\alpha\ell} \sqrt{1+ {1\over 4}\, \e^{-4\pi\alpha\ell}}
\right) \right\rbrack^2\,,
\end{equation}
while for $d=2$ we have $E_2^{\alpha,\ell}=
-\kappa^2_{\alpha,\ell}$, where $\kappa_{\alpha,\ell}$ solves the
equation
\begin{equation} \label{thres2}
\alpha + {1\over 2\pi} \left(\gamma- \ln2\right) = g_{i\kappa}(0)
\end{equation}
where $g_{i\kappa}(0)$ is given by (\ref{g2}) below -- see
\cite[Sec.~III.4]{AGHH} with an obvious correction. In both cases
$E_d^{\alpha,\ell}$ is strictly monotonous with respect to
$\alpha$. The upper edge $E_d^{\alpha,\ell}(\pi/\ell)$ of the
first spectral band is given by analogous expressions with
$g_k(0)$ replaced by $g_k(\pi/\ell)$.

Notice, in addition, that the spectrum of $\Gamma_{\alpha,Y}(k)$
for a fixed value $k\in \{\zeta:\: \im\zeta>0\} \cup \R_+$ is
absolutely continuous, because it is unitarily equivalent to an
operator of multiplication by $\alpha + {1\over 2\pi}
\left(\gamma- \ln2\right)\delta_{d,2} -g_k(\theta)$ on the
appropriate Brillouin zone, i.e. on $L^2\left(\BB_{\ell}\right)$
with $\BB_\ell:= \left(-{\pi\over\ell}, {\pi\over\ell}\right)$,
where
\begin{equation} \label{g2}
g_k(\theta) := {1\over 2\pi} \lim_{N\to\infty} \left\lbrace
\sum_{n=-N}^N {1\over 2} \left\lbrack \left(n + {\theta\ell\over
2\pi} \right)^2\! - \left( k\ell\over 2\pi\right)^2
\right\rbrack^{-1/2}\!\! - \ln N \right\rbrace.
\end{equation}
for $d=2$ and
\begin{equation} \label{g3}
g_k(\theta) := -{1\over 4\pi\ell} \ln\left\lbrack 2( \cos k\ell -
\cos\theta\ell )\right\rbrack
\end{equation}
for $d=3$. It is easy to see that the functions (\ref{g2}) and
(\ref{g3}) are nonconstant, in particular, for $k=i\kappa$ with
$\kappa>0$ they are decreasing w.r.t. $|\theta|$.

We want first to show that a deformation of the straight polymer
which satisfies the above requirement of asymptotic straightness
leaves the essential spectrum invariant.
\begin{proposition} \label{essent}
Let $Y$ satisfy the assumptions (\ref{geobound}),
(\ref{neighbor}), (a1), (a2); then $\sigma_\mathrm{ess}
(H_{\alpha,Y}) = \left\lbrack E_d^{\alpha,\ell}, E_d^{\alpha,\ell}
\left ({\pi\over\ell}\right)\right\rbrack \cup [0, \infty)$, with
the two bands overlapping for $\alpha$ large enough.
\end{proposition}
{\sl Proof:} Consider first the negative part of the spectrum. As
we have said the spectrum of $\Gamma_{\alpha,Y}(i\kappa)$ with
$\kappa>0$ for a straight polymer equals
\begin{equation} \label{str-Gamma}
\sigma\left( \Gamma_{\alpha,Y_0}(i\kappa) \right) = \left\lbrack
\alpha \!+\! {1\over 2\pi} \left(\gamma\!-\!
\ln2\right)\delta_{d,2} \!-\!g_{i\kappa}(0), \alpha \!+\! {1\over
2\pi} \left(\gamma\!-\! \ln2\right)\delta_{d,2}
\!-\!g_{i\kappa}\left({\pi\over\ell}\right) \right\rbrack.
\end{equation}
In view of Lemma~\ref{HS} below the same interval is contained in
the spectrum of $\Gamma_{\alpha,Y}(i\kappa)$, and thus by
Lemma~\ref{BS} no point of the interval
$$ I_1^{(-)}:= \left\lbrack E_d^{\alpha,\ell}, \min\left\{ 0,
E_d^{\alpha,\ell} \left ({\pi\over\ell}\right) \right\}
\right\rbrack $$
belongs to the resolvent set of the operator $H_{\alpha,Y}$, hence
$I_1^{(-)} \subset \sigma_\mathrm{ess} (H_{\alpha,Y})$. By the
same compact-perturbation argument we find that apart of a
discrete set corresponding to eigenvalues of a finite
multiplicity, the points $-\kappa^2$ outside $I_1^{(-)}$ belong to
$\rho\left(H_{\alpha,Y} \right)$, so the set $(-\infty,0)
\setminus I_1^{(-)}$ is not contained in the essential spectrum.

Let us turn to the positive halfline. Given a function $\phi\in
C_0^{\infty}([0,2])$ with $0\le \phi(r)\le 1$ and $\phi(r)=1$ for
$r\in[0,1]$, we define
$$ \psi_n(x;p,x_n):= \phi(n|x\!-\!x_n|)\, \e^{ip.x}  $$
for any $n\in\Z_0$ and $p,x_n\in\R^d$. After normalization, the
functions $\psi_n$ with an arbitrary $\{x_n\}\subset\R^2$ are
easily seen to form a Weyl sequence of the free Hamiltonian $H_0$
corresponding to the point $|p|^2$ of its essential spectrum.

Notice now that for any $N\in\Z_+$ there is a ball $B_N\subset
\R^d$ of radius $N$ which does not intersect with $Y$, for
otherwise we might take a family of such balls centered, say, at
the points $(3n_1N,0,0)$ and $(0,3n_2N,0)$ with $n_1, n_2\in\Z$,
and any array $Y$ intersecting with all of them will violate the
assumption {\em (a2)}. This means that we can choose the sequence
$\{x_n\}$ in such a way that the balls $B_{2n}(x_n)$ do not
intersect with $Y$, in which case $H_{\alpha,Y}
\psi_n(\cdot;p,x_n) =H_0\psi_n(\cdot;p,x_n)$. In this way, we have
constructed a Weyl sequence to $H_{\alpha, Y}$ for any point of
$[0,\infty)$ which completes the proof. \quad \QED \vspace{3mm}


\setcounter{equation}{0}
\section{Curvature-induced discrete spectrum} \label{disc-sec}

Now we are ready to prove the main result of this paper showing
that a non-straight array $Y$ of the class specified in
Sec.~\ref{formul} generates a nonempty discrete spectrum.
\begin{theorem} \label{dsexist}
In addition to the above assumptions, suppose that the inequality
(\ref{geobound}) is sharp for some $j,j'\in \Z$, then
$H_{\alpha,Y}$ has at least one isolated eigenvalue below
$E_d^{\alpha,\ell}$ for any $\alpha\in\R$.
\end{theorem}
{\sl Proof:} In view of Lemma~\ref{BS} we have to look for
solutions of the equation $\Gamma_{\alpha,Y}(i\kappa) \psi=\psi$,
where the operator is given by (\ref{Gamma}). We will write it as
a sum of $\Gamma_{\alpha,Y_0}(i\kappa)$ and the perturbation
$\DD_{\kappa}:= \Gamma_{\alpha,Y}(i\kappa) -
\Gamma_{\alpha,Y_0}(i\kappa)$ and investigate how the latter
affects spectral properties of a straight polymer. Since
$G_{i\kappa}(\cdot)$ is by (\ref{freeG}) strictly decreasing in
$\R_+$, the inequality (\ref{geobound}) implies
\begin{equation} \label{Dkern}
[\DD_{\kappa}]_{jj'} = \tilde g^{Y_0,i\kappa}_{jj'} - \tilde
g^{Y,i\kappa}_{jj'} = G_{i\kappa}(\ell(j\!-\!j')) -
G_{i\kappa}(y_j\!-\!y_{j'}) \le 0
\end{equation}
for any pair of non-identical $j,j'\in\Z$, while
$[\DD_{\kappa}]_{jj}=0$. We use this negativity property to show
that a sharp inequality in (\ref{geobound}) moves the threshold of
the entire spectrum.
\begin{lemma} \label{vari}
$\;\inf \sigma\left( \Gamma_{\alpha,Y}(i\kappa) \right) < \alpha
\!+\! {1\over 2\pi} \left(\gamma\!-\! \ln2\right)\delta_{d,2}
\!-\!g_{i\kappa}(0)\;$ holds for any $\kappa>0$ if the array $Y$
is not straight.
\end{lemma}
{\sl Proof:} The claim will be justified if we find $\psi\in
\ell^2(\Z)$ such that
$$ \left(\psi, \Gamma_{\alpha,Y}(i\kappa) \psi \right) < \left(
\alpha \!+\! {1\over 2\pi} \left(\gamma\!-\!
\ln2\right)\delta_{d,2} \!-\!g_{i\kappa}(0) \right) \|\psi\|^2\,,
$$
which can be in view of (\ref{Gamma}) rewritten as
\begin{eqnarray} \label{variest}
&& \sum_{j\ne j'}\, [\DD_{\kappa}]_{jj'} \bar\psi_j \psi_{j'} -
\sum_{j\ne j'}\, G_{i\kappa}(\ell(j\!-\!j'))\, \bar\psi_j
\psi_{j'} \\ && + \left( {\delta_{d,2} \over 2\pi} \ln\kappa +
{\kappa\over 4\pi} \delta_{d,3} + g_{i\kappa}(0) \right) \sum_j
|\psi_j|^2 <0 \,. \nonumber
 \end{eqnarray}
Consider first the last two terms which we shall treat by means of
the Fourier representation, $\hat\psi(\theta) = \sum_{j\in\Z}
\psi_j e_j(\theta)$, with respect to the standard trigonometric
basis $\{e_j\}$ in $L^2(\BB_{\ell})$. By Parseval relation, the
last norm equals $\int_{\BB_\ell} |\hat\psi(\theta)|^2 \D\theta$.
Furthermore, for the second term we use the Poisson summation
formula \cite{AGHH} which can be written as
$$ \sum_{j\ne 0}\, G_{i\kappa}(\ell j)\, \e^{-ij\ell\theta} =
g_{i\kappa}(\theta) + {\delta_{d,2} \over 2\pi} \ln\kappa +
{\kappa\over 4\pi} \delta_{d,3} $$
giving thus
$$ \sum_{j\ne j'}\, G_{i\kappa}(\ell(j\!-\!j'))\, \bar\psi_j
\psi_{j'} = \int_{\BB_\ell} \left( g_{i\kappa}(\theta) +
{\delta_{d,2} \over 2\pi} \ln\kappa + {\kappa\over 4\pi}
\delta_{d,3} \right) |\hat\psi(\theta)|^2 \D\theta\,. $$
Consequently, the sum of the last two terms in (\ref{variest}) is
given by the expression
\begin{equation} \label{lasttwo} \int_{\BB_\ell}
\left( g_{i\kappa}(0) \!-\! g_{i\kappa}(\theta) \right)
|\hat\psi(\theta)|^2 \D\theta \end{equation}
which is obviously non-negative. By (\ref{g2}) and (\ref{g3}) the
difference contained in the integral is a finite, smooth, and even
function on $\BB_\ell$, so there is a $c_{\kappa}>0$ such that
\begin{equation} \label{gbound}
0 \le g_{i\kappa}(0) \!-\! g_{i\kappa}(\theta) \le c_{\kappa}
\theta^2. \end{equation}
For instance, the inequality is valid with $c_{\kappa} =
{\ell\over 16\pi}\left( \sinh {\kappa\over 2}\right)^{-2}$ if
$d=3$. Let us choose now for $\psi$ the unit vector given by
$$ \psi_j = \left(\tanh \lambda\right)^{1/2} \e^{-\lambda|j|} $$
with $\lambda>0$, for which
$$ \hat\psi(\theta) = \sqrt{\ell\over 2\pi}\, \left(\tanh
\lambda\right)^{1/2} {1-\e^{-2\lambda} \over 1+\e^{-2\lambda} -
\e^{-\lambda}\, \cos \theta\ell} \,.$$
For small enough $\lambda$ we have $\tanh\lambda\le2\lambda$ and
${1\over 2} < \e^{-\lambda} < 1 - 2^{-1/2}\lambda$; then we can
estimate
$$ \hat\psi(\theta)^2 < {\ell\over 2\pi} (2\lambda)^3 \bigg(
1+\e^{-2\lambda} - \e^{-\lambda}\, \cos \theta\ell \bigg)^{-2} <
{\ell\over 2\pi} (2\lambda)^3 \left\lbrack {\lambda^2\over 2} +
{2\over\pi^2} (\theta\ell)^2 \right\rbrack^{-2}, $$
so (\ref{lasttwo}) has the following upper bound:
$$  {c_{\kappa}\over 2\pi\ell^2} (2\lambda)^3 \int_{\R} u^2
\left\lbrack {\lambda^2\over 2} + {2\over\pi^2} u^2
\right\rbrack^{-2} \D u = c_{\kappa}\pi^3\ell^{-2}\, \lambda^2\,.
$$
On the other hand, the inequality in (\ref{Dkern}) is by
assumption sharp on a non-empty subset of $\Z\times\Z$, which
means that the first term in (\ref{variest}) is
$$ \sum_{j\ne j'}\, [\DD_{\kappa}]_{jj'}\, \tanh\lambda\;
\e^{-\lambda(|j|+|j'|)} \le -c\lambda $$
for some positive $c$ and all $\lambda$ small enough. Hence this
term dominates the \lhs of (\ref{variest}) as $\lambda\to 0+$;
this result yields the sought trial vector. \quad \QED
\vspace{2mm}

Our next goal is to show that the perturbation (\ref{Dkern}) is a
compact operator provided the array $Y$ is sufficiently straight
at large distances.
\begin{lemma} \label{HS}
Under the assumption (a2), $\DD_{\kappa}\,$ is Hilbert-Schmidt if
$\mu>\,{1\over 2}\,$.
\end{lemma}
{\sl Proof:} We have to estimate the \rhs of (\ref{Dkern}). For
brevity, we introduce the following notation for arguments
appearing at this expression,
$$ \varrho \equiv \varrho_{jj'} := |y_j\!-\!y_{j'}|\,, \quad
\sigma \equiv \sigma_{jj'} := \ell|j\!-\!j'|\,. $$
We employ the fact that the free Green's functions (\ref{freeG})
are concave for $k=i\kappa$. This yields
\begin{equation} \label{Dest}
0 \le G_{i\kappa}(\varrho) - G_{i\kappa}(\sigma) \le -\,\varrho\,
G'_{i\kappa}(\varrho)\: {\sigma\!-\!\varrho \over \varrho} \,,
\end{equation}
where the derivative at the \rhs is equal to $-\kappa
K_1(\kappa|\cdot|)$ for $d=2$, and $-{\e^{-\kappa|\cdot|}\over
4\pi|\cdot|^2} (1\!+\!\kappa|\cdot|)$ for $d=3$. Notice that the
function $\varrho \mapsto \varrho\, G'_{i\kappa}(\varrho)$ is
bounded in $(0,\infty)$ for $d=2$; for $d=3$ it has a singularity
at the origin but it is bounded in $[c_1\ell, \infty)$, \ie, for
all values of $\varrho$ giving a non-vanishing contribution to
$\DD_{\kappa}$. At the same time we have
\begin{equation} \label{geo}
0\le {\sigma\!-\!\varrho \over \varrho} \le {1\!-\!c_1 \over c_1}
\end{equation}
in view of $c_1\sigma\le \varrho\le\sigma$, hence the matrix
elements of $\DD_{\kappa}$ are bounded.

This is not enough, of course, we need to know their decay
properties. Away of the sector $S_\omega$ we employ the fact that
there is a $c>0$ such that
\begin{equation} \label{Kest}
-\,\varrho\, G'_{i\kappa}(\varrho) \le c\, e^{-\kappa\varrho/2}
\le c\, e^{-c_1\kappa\sigma/2}
\end{equation}
holds for all non-zero $\varrho, \sigma$, while in the said sector
we have by {\em (a2)} the estimate
\begin{equation} \label{geo2}
{\sigma\!-\!\varrho \over \varrho} \le {\sigma\!-\!\varrho \over
c_1\sigma} \le {c_2\over c_1}\, \left\lbrack 1+|j\!+\!j'|^{2\mu}
\right\rbrack^{-1/2}\,. \end{equation}
Combining the estimates (\ref{Dest})--(\ref{geo2}) we get a bound
to the Hilbert-Schmidt norm in question:
\begin{equation} \label{HSnorm}
\sum_{j\ne j'}\, [\DD_{\kappa}]_{jj'}^2 \le \left(1\!-\!c_1 \over
c_1\right)^2\, c^2 \sum_{\Z^2 \setminus S_{\omega}}
e^{-c_1\kappa|j-j'|}  +\, \left(c c_2\over c_1\right)^2\:
\sum_{S_{\omega}} {e^{-c_1\kappa|s-s'|} \over 1+|s+s'|^{2\mu}}\,.
\end{equation}
Denoting $s=j'\!-\!j$, we can rewrite the first sum at the \rhs as
$$ 2\, \sum_{s=1}^{\infty}\, \sum_{j= \left\lbrack - {s\over
1-\omega} \right\rbrack +1}^{\left\lbrack {s\omega\over 1-\omega}
\right\rbrack}\, e^{-c_1\kappa s} \le 2\, \sum_{s=1}^{\infty}\,
\left( 1+ s\,{1+\omega\over 1-\omega} \right)\, e^{-c_1\kappa s}
$$
with the last series obviously convergent. The second series at
the \rhs of (\ref{HSnorm}) is not diminished if we sum over all
the $\Z^2$. Passing then to $j\pm j'$ as the new summation
variables we see that it is finite for $\mu>\, {1\over 2}\,$.
\quad \QED \vspace{2mm}

As the last ingredient, we need the following continuity result:
\begin{lemma} \label{cont}
Under the same assumptions as above, the map $\;\kappa\mapsto
\Gamma_{\alpha,Y}(i\kappa)\,$ is operator-norm continuous and
$\,\inf \sigma\left(\Gamma_{\alpha,Y}(i\kappa) \right)\to
\pm\infty $ as $\kappa\to\infty$ and $\kappa\to0+$, respectively.
\end{lemma}
{\sl Proof:} The claim holds for the ``free'' operator
$\Gamma_{\alpha,Y_0}(i\kappa)$, because the functions
$\kappa\mapsto g_{i\kappa}(\theta)$ are continuous for any
$\theta\in \BB_\ell\,$, and $g_{i\kappa}(0)$ which determines the
spectrum bottom has the needed limits. More specifically,
$g_{i\kappa}(0) \to-\infty$ as $\kappa\to\infty$, its asymptotics
being logarithmic for $d=2$ and linear for $d=3$, while
$g_{i\kappa}(0) \to+\infty$ as $\kappa\to0+$. It is thus
sufficient to check that the map $\kappa\mapsto \DD_{\kappa}$ is
continuous and remains bounded as $\kappa\to\infty$. We employ the
inequality
$$ [\DD_{\kappa}\!-\!\DD_{\kappa'}]_{jj'}^2 \le 2\left(
[\DD_{\kappa}]_{jj'}^2 +[\DD_{\kappa'}]_{jj'}^2 \right) \le
4[\DD_{\kappa_0}]_{jj'}^2  $$
which holds true for any $\kappa_0 \le \min(\kappa,\kappa')$.
Hence the series expressing the HS-norm of $\DD_{\kappa}\!-\!
\DD_{\kappa'}$ can be uniformly majorized and the limit may be
interchanged with the sum giving
$$ \| \DD_{\kappa}\!-\!\DD_{\kappa'} \|_\mathrm{HS} \to 0 \qquad
\mathrm{as} \qquad \kappa'\to\kappa\,.  $$
At the same time, the estimate (\ref{HSnorm}) shows that not only
the norm remains bounded, but even $ \| \DD_{\kappa}
\|_\mathrm{HS} \to 0$ as $\kappa\to\infty$ which concludes the
proof. \quad \QED \vspace{2mm}

\noindent {\sl Proof of Theorem~\ref{dsexist}, continued:} We have
$\inf \sigma\left( \Gamma_{\alpha,Y}(i\kappa) \right) < \inf
\sigma\left( \Gamma_{\alpha,Y_0}(i\kappa) \right)$ by
Lemma~\ref{vari} whenever the array $Y$ is not straight. On the
other hand, a deformation $Y$ of $Y_0$ which is asymptotically
straight in the sense of assumption {\em (a2)} leaves by
Lemma~\ref{HS} the essential spectrum invariant, so the part of
$\sigma\left( \Gamma_{\alpha,Y}(i\kappa) \right)$ outside the
interval $\ref{str-Gamma}$ consists of isolated eigenvalues of a
finite multiplicity at most. Putting the two results together, we
find that for any $\kappa>0$ there is at least one such eigenvalue
$\lambda_{\alpha,Y}(\kappa)$ below $\inf \sigma\left(
\Gamma_{\alpha,Y_0}(i\kappa) \right)$. Finally, by
Lemma~\ref{cont} the function $\lambda_{\alpha,Y}(\cdot)$ is
continuous and $\lambda_{\alpha,Y}(\kappa)\to \pm\infty$ as
$\kappa\to\infty$ and $\kappa\to0+$, respectively. This means that
for any $\alpha\in\R$ there is a point $\kappa_0$ such that
$\lambda_{\alpha,Y}(\kappa_0)=0$, and the eigenvalue $-\kappa_0^2$
which corresponds to it by Lemma~\ref{BS} satisfies $-\kappa_0^2 <
E_d^{\alpha,\ell}$. \quad \QED


\setcounter{equation}{0}
\section{Concluding remarks} \label{card}

Having established the existence of ``trapped modes'' in locally
curved polymers, we are naturally interested on the cardinality of
$\sigma_\mathrm{disc}(H_{\alpha,Y})$, or in other words, how rich
the discrete part of the spectrum could be. We want to illustrate
that the number of bound states can be any positive integer
provided $Y$ is curved ``enough''.
\begin{example}
{\rm Consider the set $Y_{M,N}$ obtained by an appropriate
ordering of $\{ (j,j'):\: j\ne j',\: \max(j,j')\le M\} \cup \{
(j,j'):\: |j|\le N\,\}$, to which we ascribe the operator $H_{M,N}
\equiv H_{\alpha,Y_{M,N}}$, again with the same $\alpha\in\R$ at
all points. Depending on $\alpha$, it has at most $2M(2M\!+\!1)+
2N +1$ eigenvalues naturally ordered as $\mu^{(1)}_{M,N} \le
\mu^{(2)}_{M,N} \le \dots <0$.

Using the monotonicity of point-interaction Hamiltonians w.r.t.
the coupling constants as in \cite[Sec.~II.1.1]{AGHH}, one checks
that above operator family is non-increasing w.r.t. both $M$ and
$N$ which means, in particular,
$$ \mu^{(j)}_{M,N} \ge \max\left( \mu^{(j)}_{M+1,N},
\mu^{(j)}_{M,N+1} \right) $$
for each $j, M, N$. The sequence $\{H_{M,M}\}_{M=1}^{\infty}$
converges in the strong resolvent sense respectively to the
Hamiltonian of the square crystal if $d=2$, and of a monomolecular
layer if $d=3$ \cite[Secs.~III.1.6, III.4]{AGHH}. Both have
absolutely continuous spectra with the threshold strictly below
the number $E_d^{\alpha,\ell}$ described in Sec.~\ref{ess-sec}.
Hence to any positive integer $j_0$ one can find $M \equiv M(j_0)$
such that $H_{M,M}$ has $j_0$ eigenvalues less than
$E_d^{\alpha,\ell}$.

On the other hand, the sequence $\{H_{M,N}\}_{N=M}^{\infty}$
converges in the strong resolvent sense to the operator
$H_{\alpha,Y}$ corresponding to the array $Y \equiv Y_{M,\infty}$
which is straight except for the central part where it is
``tightly packed''. Using the monotonicity again we find that
$H_{\alpha,Y}$ has at least $j_0$ eigenvalues below the
essential-spectrum threshold $E_d^{\alpha,\ell}$. }
\end{example}

Let us conclude the paper with several remarks. One may ask about
the meaning of the asymptotic straightness requirement {\em (a2)}.
In a companion paper \cite{EI} which treats an analogous problem
for measure perturbations of two-dimensional Schr\"odinger
operators supported by curves we show that a sufficient condition
for a planar curve to satisfy such a condition is that its signed
curvature decay with respect to the arc length $s$ is $o\left(
|s|^{-5/4 -\epsilon}\right)$.

Another natural question is what a curvature will do to the
spectrum of a two-dimensional array under influence of a constant
magnetic field, where in the straight case we have absolutely
continuous bands sandwiched between the Landau levels \cite{EJK}.
The argument of the present paper does not extend to this
situation, because its crucial ingredient is the real-valuedness
and monotonicity of the free Green's function which is no longer
true for magnetic Schr\"odinger operators. Thus the question
deserves a separate treatment which we postpone to another paper.
The same is true is true for the continuous-spectrum part of the
non-magnetic problem where the curvature leads in general to a
non-trivial scattering.


\subsection*{Acknowledgments}

The research has been partially supported by GAAS project \#
1048801.


\begin{thebibliography}{99}
\bibitem[AGHH]{AGHH}
S.~Albeverio, F.~Gesztesy, R.~H\o egh-Krohn, H.~Holden: {\em
Solvable Models in Quantum Mechanics}, Springer, Heidelberg 1988.
\vspace{-1.8ex}
\bibitem[DE]{DE}
P.~Duclos, P.~Exner: Curvature-induced bound states in quantum
wave\-guides in two and three dimensions, {\em Rev.Math.Phys.}{\bf
7} (1995), 73-102. \vspace{-1.8ex}
 \bibitem[EI]{EI}
P.~Exner, T.~Ichinose: Geometrically induced spectrum in curved
leaky wires, {\em in preparation} \vspace{-1.8ex}
 \bibitem[EJK]{EJK}
P.~Exner, A.~Joye, H.~Kova\v{r}\'{\i}k: Edge currents in the
absence of edges, {\em Phys. Lett.} {\bf A264} (1999), 124-130.
\vspace{-1.8ex}
 \bibitem[E\v S]{ES}
P.~Exner, P.~\v{S}eba: Bound states in curved quantum waveguides,
{\em J.Math. Phys.} {\bf 30} (1989), 2574-2580. \vspace{-1.8ex}
 \bibitem[Hu]{Hu}
N.E. Hurt: {\em Mathematical Physics of Quantum Wires and
Devices}, Kluwer, Dordrecht 2000. \vspace{-1.8ex}
 \bibitem[LCM]{LCM}
J.T. Londergan, J.P. Carini, D.P. Murdock: {\em Binding and
scattering in two-dimensional systems}, LN~m60, Springer, Berlin
1999. \vspace{-1.8ex}
\end{thebibliography}
\end{document}